\documentclass[12pt,pra,unsortedaddress,superscriptaddress,a4paper,nofootinbib,%
nobalancelastpage,preprintnumbers,showpacs,tightenlines,eqsecnum]{revtex4}
\usepackage{amsmath, amsbsy}
\usepackage{bbm}
\usepackage[pdftex]{graphicx}
\usepackage{aas_macros}
\def\vec#1{\mathbf{#1}}
\def\be{\begin{equation}}
\def\bel#1{\begin{equation}\label{#1}}
\def\ee{\end{equation}}
\def\d{\mathrm{d}}

\def\H{\ensuremath{\mathrm{H}}}
\def\Hm{\ensuremath{\mathrm{H}^-}}

\def\Psm{\ensuremath{\mathrm{Ps}^-}}

\def\Hdp{\ensuremath{\mathrm{H}_2{}^+}}
\begin{document}
\title{The domain of stability\\ of unnatural parity states of three unit charges}
\author{J.-M. Richard}
\email{jean-marc.richard@lpsc.in2p3.fr}
\affiliation{Laboratoire de Physique Subatomique et de Cosmologie,
 IN2P3-CNRS, Universit\'e Joseph Fourier,  INPG,
 53, avenue des Martyrs, Grenoble, France}

\begin{abstract}
We investigate the domain of masses for which a state with total orbital momentum $L=1$ and unnatural parity $P=+1$ exists for the Coulomb systems $(m_1^+,m_2^-,m_3^-)$.
\end{abstract}

\date{\today}

\maketitle
\section{Introduction}
There is an abundant literature on Coulomb systems, in particular,  ions made of three unit charges,  $(m_1^\pm,m_2^\mp,m_3^\mp)$. For the ground state with total angular momentum and parity $L^P=0^+$, some particular configurations have been studied in great detail, and the map of the stability domain has been drawn, when the masses $m_i$ are varied. For a review, and references, see, e.g., \cite{2005PhR...413....1A}. 

The study of this $0^+$ ground state indicates a number of interesting features.  There is a variety of situations: deep binding  as for $\Hdp$, instability as for $(p,\bar{p},e^-)$, and states at the edge of binding, such as $\Psm (e^+,e^-,e^-)$ or $\Hm (p,e^-,e^-)$. 
The stability is reinforced when two particles become identical ($m_2= m_3$). Schematically, the system has two components: a $(m_1^+,m_2^-)$ cluster with a loosely attached third particle, or $(m_1^+,m_3^-)$ cluster weakly linked to the second particle; when the thresholds become identical, the two components have maximal interference to build a  compound system.

For a fragile system such as \Hm, the Hartree--Fock picture fails. No variational wave function $f(r_{12}) f(r_{13})$ gives an expectation value lower than the threshold energy. One needs either a explicit $r_{23}$-dependence in the wave function, or a breaking of factorization, as in the famous Chandrasekhar wave function \cite{1944ApJ...100..176C},
\be\label{eq:chan}
\Psi=\exp(-a \,r_{12}-b\,r_{13})+\exp(-b \,r_{12}-a\,r_{13})~,
\ee
in which the particle identity is first broken and then restored by a counterterm, a strategy sometimes named ``unrestricted Hartree--Fock'' \cite{PhysRev.172.7}. 
In natural units, the $\H(1s)+e^-$ threshold for $\Hm$ is at $-0.5$, the above wave-function gives  an energy of about $-0.513$, and the elaborate calculations about $-0.528$. So it is not surprising that there is only one bound state below this lowest threshold, as rigorously demonstrated by Hill \cite{hill:2316}.

However, the attention was later paid on the positive-parity state where each electron is a $p$-wave, coupled to  form a total angular momentum $L=1$.
Such  a state with $L^P=1^+$ cannot decay into $\H(1s)+e^-$ as long as radiative corrections and spin effects are neglected. Its effective threshold consists of $\H(2p)+e^-$, at $-0.125$. Variational calculations, first at Oslo \cite{Hmoneplus} and confirmed elsewhere (see, e.g., \cite{PhysRevLett.24.126,1979JChPh..71.4611J}), give an energy of about $-0.12535$, just below the threshold. Grosse and Pittner~\cite{grosse:1142} have shown that this is the only unnatural-parity state of \Hm\ with this type of stability.  But higher configurations of unnatural parity exist, for instance by attaching a positron to a two-electron system of unnatural parity 
\cite{bromley:062505}.

Starting from  \Hm\  and exchanging the masses leads to \Hdp, whose ground state is deeply bound and excitation spectrum very rich, and not surprisingly, includes a stable $1^+$ state. However, Mills \cite{PhysRevA.24.3242} investigated the case of equal masses, i.e., \Psm, and  found the $1^+$ state to be unbound. More precisely, he studied the  $1^+$ state for any configuration  $(M^+,m^-,m^-)$ and got stability everywhere except for 
\be\label{eq:Mills}
0.059 \lesssim m/M \lesssim 2.36~.
\ee  
This estimate was confirmed by Bhatia and Drachman \cite{PhysRevA.28.2523}. 
It is the aim of the present note to check the domain of stability for $(M^+,m^-,m^-)$ found in \cite{PhysRevA.24.3242,PhysRevA.28.2523} (without trying to challenge their accuracy), and to extend the study to the case of unequal masses for the negative charges. It is organized as follows. In Sec.~\ref{se:prop}, we summarize the rigorous results on the stability domain. The variational method and the results are presented in Sec.~\ref{se:var}, and 
Sec.~\ref{se:out} is devoted to some conclusions. 
\section{Properties of the stability domain}\label{se:prop}
\begin{figure}
\begin{minipage}{.5\textwidth}
\begin{center}
\includegraphics[width=.8\textwidth]{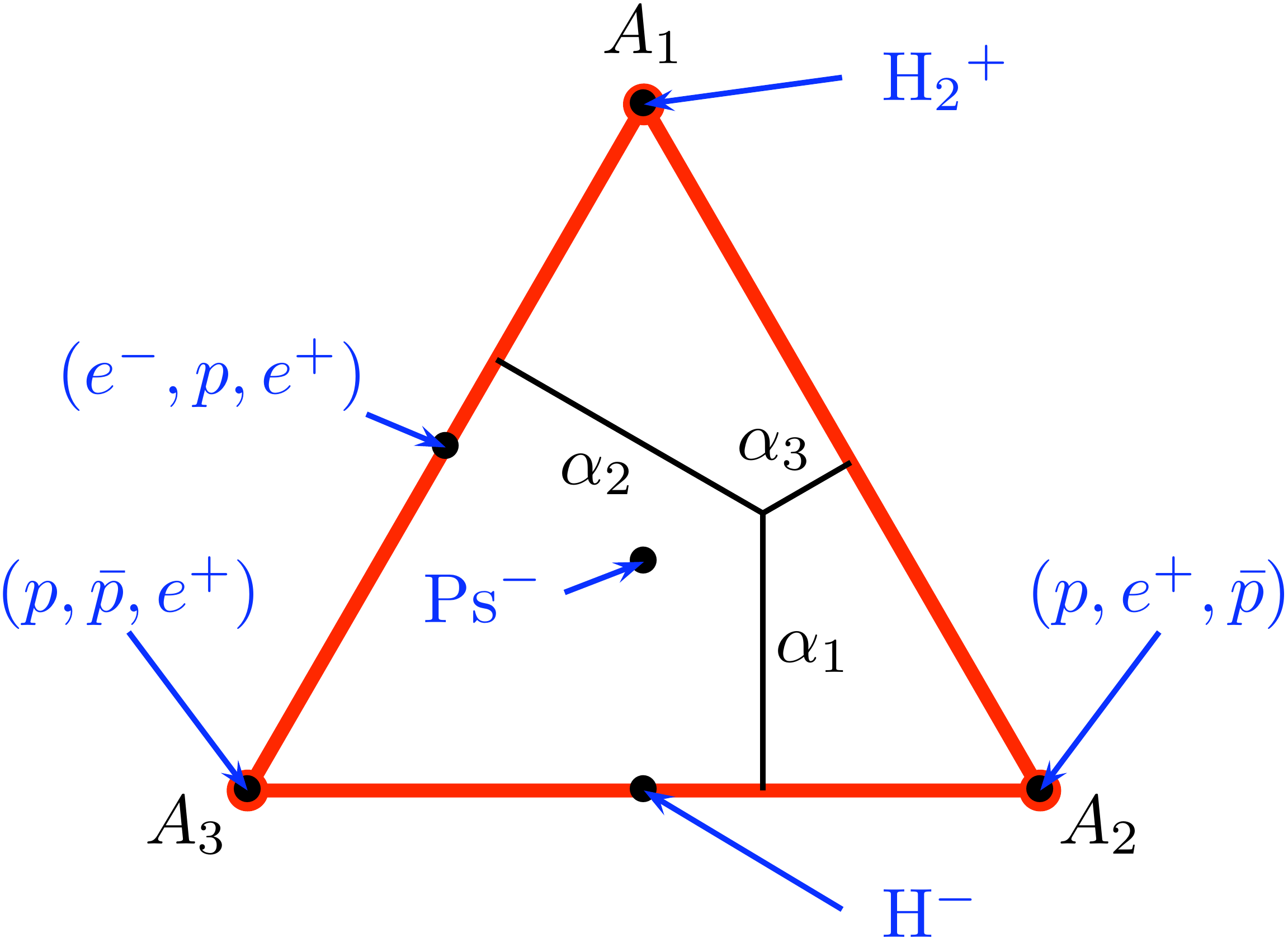}
\end{center}
\end{minipage}
\begin{minipage}{.4\textwidth}
\caption{\label{fig:Dal} Representation of a $(m_1^+,m_2^+,m_3^+)$ Coulomb system from its normalized inverse masses $\alpha_i=m_i^{-1}/(m_1^{-1}+m_2^{-1}+m_3^{-1})$. Some special cases are also shown.}
\end{minipage}
\end{figure}

We consider the Hamiltonian
\begin{equation}\label{Hamiltonian}
H=\frac{\vec{p}_1^2}{2 m_1}+\frac{\vec{p}_2^2}{2 m_2}+\frac{\vec{p}_3^2}{2 m_3}-
\frac{1}{r_{12}}-\frac{1}{r_{23}}+\frac{1}{r_{23}}~, 
\end{equation} 
and focus on the lowest $L^P=1^+$ state. By scaling, each charge can be set to unity, and one can impose that the inverse masses $\alpha_i=1/m_i$ obey $\alpha_1+\alpha_2+\alpha_3=1$. Thus, as done for the $0^+$ ground state \cite{PhysRevA.46.3697,2005PhR...413....1A}, each physical system with $m_i>0$ can be represented as a point inside an equilateral triangle, and the inverse masses $\alpha_i$ are proportional to the distances to the sides. See Fig.~\ref{fig:Dal}.

In this representation, the frontier between stability and instability has the following properties:
\begin{itemize}
\item It is symmetric with respect to the vertical axis, due to $2\leftrightarrow3$ exchange, 
\item In each of the half-triangles limited by the symmetry axis, the instability domain is convex.
\item The instability domain including, say, the point $A_2$ with inverse masses $(0,1,0)$  is star-shaped with respect to $A_2$. (A semi-straight line starting from $A_2$ enters at the most once a stability domain until it reaches the symmetry axis.)
\item If an energy 
\be \label{eq:def-eps}
E=E_\text{th}\,(1+\epsilon)~,
\ee
is found for  $(M^+,m^-,m^-)$, where $E_\text{th}=-M\,m/[2(M+m)]$ is the $(M^+,m^-)$ threshold, and $\epsilon$ measures the relative fraction of extra binding, and  if one considers an asymmetric combination $(m_1^+,m_2^-,m_3^-)$ with $m_1=M$ and the same average inverse mass in the negative sector, i.e., $m_1^{-1}+m_2^{-1}=2 m^{-1}$, then stability extends at least up to 
\be\label{eq:min-wid}
\left| m_1^{-1}-m_2^{-1} \right| = \frac{2\,\epsilon}{1+\epsilon}\left(M^{-1}+m^{-1}\right)~.
\ee
\end{itemize}
The proof is the same as for the $0^+$ ground state \cite{PhysRevA.46.3697,2005PhR...413....1A}. However, in this latter case, its was also demonstrated that any configuration with $m_2=m_3$ has at most one bound state \cite{hill:2316}, so that the entire stability domain was connected, and clustered near the symmetry axis. This is not the case of $L^P=1^+$, as no stable state exists for equal masses \cite{PhysRevA.24.3242,PhysRevA.28.2523}. Hence, the stability domain includes two separate islands, one around \Hdp, and another around \Hm.
\section{Variational method and results}\label{se:var}
\subsection{Trial wave function}
For scalar states, a generalization of (\ref{eq:chan}) is 
\be\label{eq:scalar}
\Psi=\sum_i \gamma_i\,\exp(-a\,x - b\,y - c\, z)~,
\ee
where $\vec{x}=\vec{r}_{23}$, 
\dots join the particles and $x=|\vec{x}|$, \dots are the relative distances. After integration over the trivial angular variables, one is left with integrating over $x\,y\,z\,\d x\, \d y \, \d z$, submitted to the triangular inequality. The matrix elements can be deduced from the generic function
\begin{equation}\label{eq:generic}
F(\alpha, \beta,\gamma)=\iiint\limits_{|x-y|\le z \le x+y}  \exp(-\alpha\, x -\beta\, y - \gamma\, z)\, \d x\,\d y\,\d z=\frac{4}{(\alpha+\beta)(\beta+\gamma)(\gamma+\alpha)}~,
\end{equation}
and its derivatives.

For a $L^P=1^+$ state with projection $L_3=j$, a trial function is the superposition
\be\label{eq:axial}
\Psi=(\vec{y}\times \vec{z})_j\,\sum_i \gamma_i\, \exp(- a_i\, x - b_i \,y - c_i\, z)~,
\ee 
 with the possibility of enforcing explicitly the ${x\leftrightarrow y}$ symmetry if $m_2=m_3$.  Once the integration is carried out over the angular variables, one can express, again, all the matrix elements in terms of the derivatives of $F$. For a given set of range parameters, the weights $\gamma_i$ are given by a generalized eigenvalue problem. It is possible to assign the $a_i,\, b_i,\, c_i$  to have values in a restricted set $(u, v, \ldots)$, i.e., $(a_i,\, b_i,\, c_i)= (u,u,u), (u,u,v), \ldots (u,v,w), \dots (w,w,w), \ldots$, and even assume $(u,v,\ldots)$ in a kind of progression, in order to simplify the minimization over the non-linear parameters, without a significant loss of accuracy. This is similar to the strategy used, e.g., by Kamimura et al.\ when handling   the expansions  over Gaussians \cite{2003PrPNP..51..223H}.
\subsection{\Hm-like states}
For \Hm, the energy is found to be $E\simeq-0.1253$, as in the literature. For other symmetric configurations $(M^+,m^-,m^-)$, we obtain stability for $M/m \gtrsim 2.39$, in good agreement with Mills's estimate (\ref{eq:Mills}). The width of the stability domain is very narrow. For an infinitely massive proton, we find stability only for $m_3/m_2\lesssim 1.006$ (if $m_3\ge m_2$). This is very close to the estimate based on (\ref{eq:min-wid}), as expected for such a small width. For instance, the $1^+$ state does not exist for the very exotic $(p,\pi^-,\mu^-)$ system.
\subsection{\Hdp-like states}
Stability is found for $(M^+,m^-,m^-)$ with $M/m\lesssim 0.055$, again in good agreement with Eq.~(\ref{eq:Mills}). On the side of the triangles, we obtain stability for the inverse masses $(1-\alpha,\alpha, 0)$ and the symmetric  points $(1-\alpha,0,\alpha)$ for $0\le\alpha\lesssim 0.094$.  For the point of the symmetry axis with the  inverse masses $(1-\alpha,\alpha/2,\alpha/3)$, the relative excess of energy is found to be $\epsilon\simeq 0.034$, and Eq.~(\ref{eq:min-wid}) is underestimates the width of the domain by about $15\%$.
\begin{figure}[!htbc]
\begin{center}
\includegraphics[width=.45\textwidth]{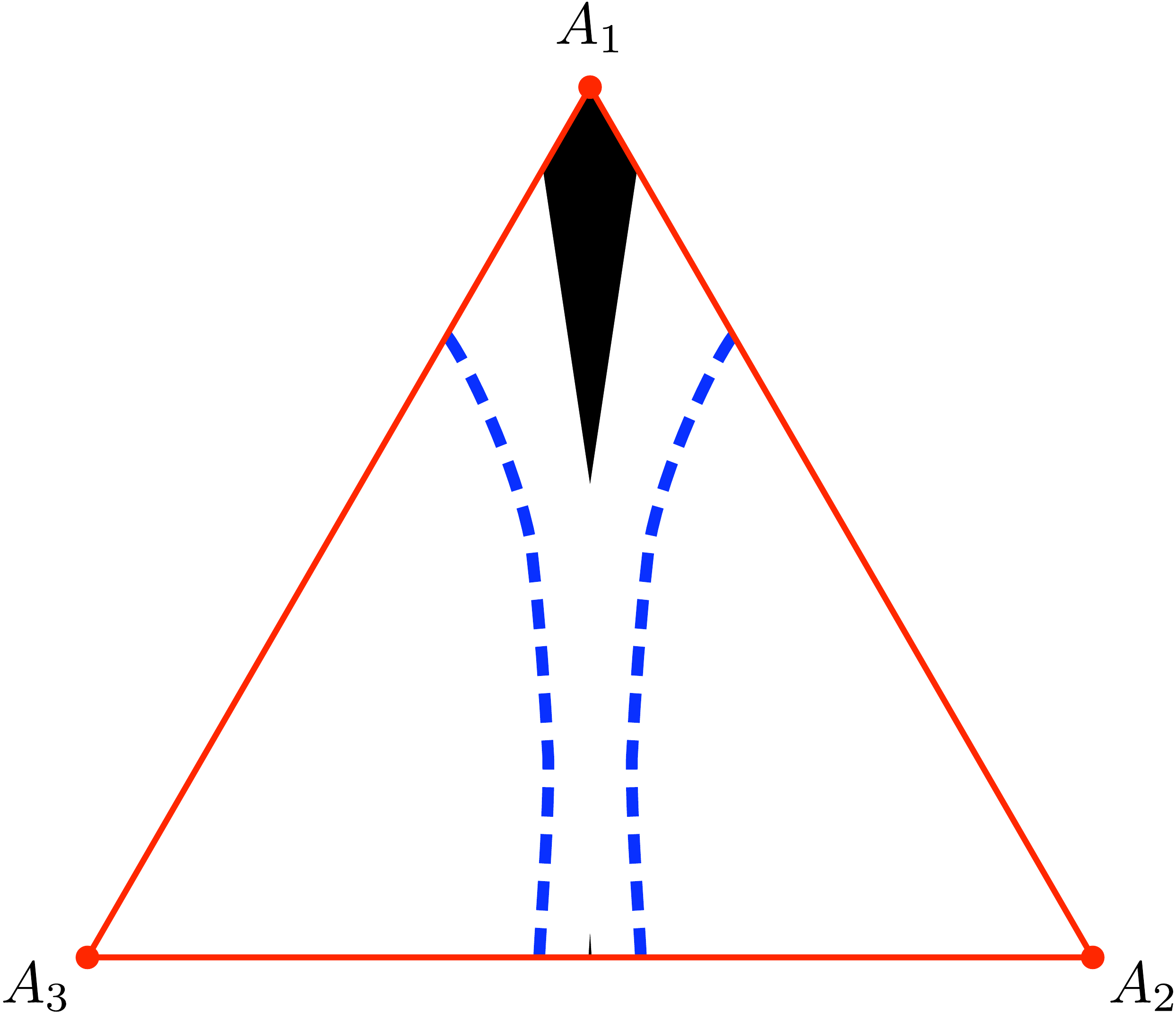}\qquad
\raisebox{.3cm}{\includegraphics[width=.45\textwidth]{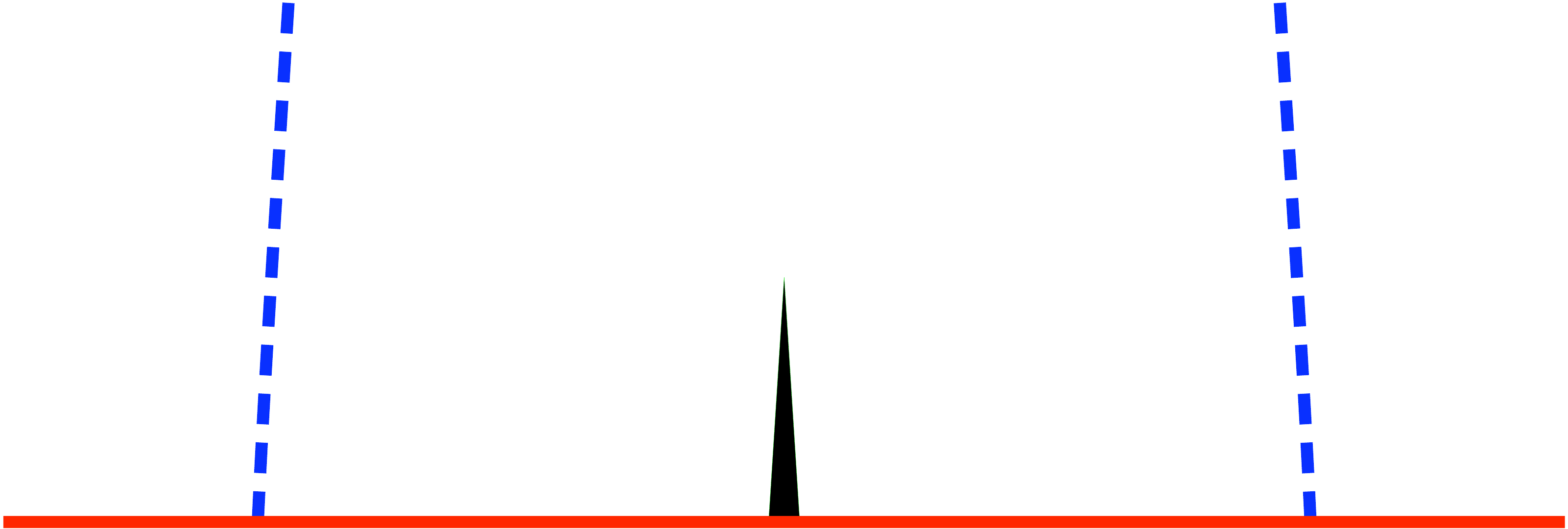}}
\end{center}
\caption{\label{fig:recap} Domain of stability for $(m_1^+,m_2^-,m_3^-)$ in the triangle of normalized inverse masses: the ground state $0^+$ (dashed lines) and the unnatural-parity state $1^+$ (solid). On the right, magnification of the bottom part of the plot.}
\end{figure}
\subsection{Summary}
The domain of stability contains at least  the areas displayed in Fig.~\ref{fig:recap}. The frontier is almost straight, so the convexity of the instability domain is weakly pronounced. 
The spike around \Hm\ is very small and extremely narrow, and shows up only after magnification. For comparison, the domain of stability of the ground state $0^+$ is also shown.

\section{Outlook}\label{se:out}
The case of unnatural-parity states of three unit-charge particles displays an interesting -- and rare -- example of \emph{discontinuous} stability domain, where stability disappears and shows up again, when some constituent masses are varied continuously and monotonously. This means that the transition from a molecular type of binding (\Hdp) to a halo-type of binding (\Hm) involves some more fragile intermediate dynamics of stability.

\begin{acknowledgments}
Discussions with M.~Asghar,  A.~Czarnecki, H.~H{\o}gaasen, A.~Martin, J.~MItroy, P.~Sorba and Tai T.~Wu are gratefully acknowledged. 
\end{acknowledgments}


\end{document}